\documentclass[11pt]{article}

\usepackage{amsfonts,amssymb,cancel,chet,dsfont,graphicx,mathrsfs,multirow,pgfplots,subfig,tikz}

\usetikzlibrary{external}
\title{Magnetar Hard X-Ray Emission\\from Axion-like Particle Conversion}

\author{Jean-Fran\c{c}ois Fortin\email{jean-francois.fortin@phy.ulaval.ca} and Marianne Gratton\email{marianne.gratton.1@ulaval.ca}}

\affiliation{
D\'epartement de Physique, de G\'enie Physique et d'Optique,\\Universit\'e Laval, Qu\'ebec, QC G1V 0A6, Canada
}

\abstract{%
We explore the possibility that axion-like-particles (ALPs), which would be produced in the core of magnetars and would then convert in the magnetosphere into photons, can explain magnetar hard X-ray spectra.  We remark that this scenario would also provide answers to some questions related to magnetar heating.  Indeed, considering that magnetars have: 1) hard X-ray spectra that are difficult to explain with known mechanisms; 2) large photon luminosities that force high core temperatures; 3) high core temperatures that imply large neutrino emissivities; 4) and large neutrino emissivities that lead to small magnetar lifetimes in contradiction to observations---explaining the hard X-ray spectra with ALPs could decrease the core temperatures and thus the neutrino emissivities, allowing for longer magnetar lifetimes as expected from observations.  In this work, we initiate the study of this scenario for three magnetars with extreme luminosities, and conclude that the general idea is likely worth investigating in more detail.
}

\date{June 2022} 

\begin{document}

\maketitle



\section{Introduction}\label{SIntro}

Magnetars are neutron stars with extreme magnetic fields in excess of the critical magnetic field $B_c=m_e^2/\sqrt{4\pi\alpha}=4.414\times10^{13}\,\text{G}$ \cite{Turolla:2015mwa,Beloborodov:2016mmx,Kaspi:2017fwg}.  Here $m_e$ is the electron mass and $\alpha$ is the fine structure constant.  Such stars lose their heat energy through neutrino emission and surface radiation.  Analysis of magnetar soft X-ray spectra suggests surface temperatures $T_s\gtrsim10^6\,\text{K}$ which lead to enormous losses of energy through neutrino emission \cite{Beloborodov:2016mmx}.  This energy sink is however too large to allow magnetars with ages in the $1-10\,\text{kyr}$ range, in contradiction to observations \cite{Beloborodov:2016mmx,Vigano:2013lea} (from now on, this problem is referred to as the magnetar age/temperature problem).  Moreover, although there exist several proposals \cite{Thompson:2004yg,Heyl:2005an,Beloborodov:2006qh,Baring:2006hi,Lyubarsky:2007gj,Beloborodov:2012ug},\footnote{All such mechanisms rely on the extreme magnetic field of magnetars: to accelerate charged particles leading to synchroton radiation, to upscatter photons through resonant magnetic Compton scattering, \textit{etc.}} there is still the possibility that another mechanism could contribute to the hard X-ray spectra that magnetars exhibit.

Axions, and more generally axion-like particles (ALPs), are extremely light and weakly-coupled hypothetical particles that have been proposed to solve several issues within the Standard Model (SM) of particle physics \cite{Peccei:1977hh,Peccei:1977ur,Wilczek:1977pj,Weinberg:1977ma}.  In their first implementation, ALPs were introduced to explain the lack of CP violation in strong interactions through the Peccei-Quinn mechanism.  For appropriate mass and coupling, ALPs could also play a role in dark matter \cite{Preskill:1982cy,Abbott:1982af,Dine:1982ah}, solving two outstanding problems of the SM at once.  ALPs are however very hard to search for since they interact so weakly with SM particles.

If they exist, ALPs could play an important role in the dynamics of magnetars.  Indeed, as neutrinos, ALPs would be copiously produced in such dense and hot environments \cite{Iwamoto:1984ir,Nakagawa:1987pga,Nakagawa:1988rhp,Iwamoto:1992jp,Raffelt:1996wa,Umeda:1997da,Maruyama:2017xzl,Paul:2018msp}, exiting the star unhindered and producing an extra energy sink.  This additional loss of energy would at first seem to imply that ALPs would exacerbate the magnetar age/temperature problem mentioned above.  However, due to their coupling to photons, ALPs can oscillate efficiently into photons in the magnetosphere, leading to a source of soft and hard X-ray photons \cite{Raffelt:1987im,Lai:2006af,Chelouche:2008ta,Jimenez:2011pg,Perna:2012wn,Marsh:2015xka,Graham:2015ouw,Fortin:2018ehg,Fortin:2018aom}.  Hence, ALPs emission could possibly explain (part of) the magnetar hard X-ray spectra, thus leading to smaller surface and core temperatures, and therefore help alleviate the magnetar age/temperature problem.  Hence, from this mechanism, ALPs could help solve another issue, not in the SM, but with respect to magnetar heating.

In this paper we investigate this idea in some detail.  We consider three magnetars with large X-ray luminosities, and estimate their core temperatures in two different manners: indirectly through their surface temperatures and more directly through their X-ray luminosities (Section~\ref{SMag}).  We then turn on ALP couplings to compute ALP production and the associated X-ray luminosities from ALP-to-photon conversion (Section~\ref{SALPs}).  With this result, we perform fits to the observed spectra and determine the best core temperatures and ALP coupling constants for each magnetar (Section~\ref{SResults}).  Our simple analysis demonstrates that core temperatures with ALPs emission can be smaller than core temperatures without ALPs emission by up to one order of magnitude (depending on the method used to estimate the core temperature), leading to a substantial drop in neutrino emission, and thus, an increase in allowed magnetar lifetimes in agreement with observations (although the best-fit ALP coupling constants are in tension with other bounds).  Finally, we find confidence-level contours in the core temperature-axion coupling constants plane to compare the suggested ALPs coupling constants from different magnetar fits.  For different magnetar radii, it is possible to obtain a partial agreement for the axion coupling constants---which would technically be fixed for all magnetars---lending some credence to our scenario (Section~\ref{SConclusion}).

We stress that our analysis is rudimentary, with several approximations.  For example: only neutrons are considered, the effects of protons are discarded; the one-pion-exchange approximation \cite{Friman:1979ecl,Brinkmann:1988vi,Machleidt:2017vls} is used instead of the soft-radiation approximation \cite{Hanhart:2000ae}; only the most dominant channels are considered, no exotic processes (superfluidity, \textit{etc.}) are included \cite{Yakovlev:1999sk,Page:2013hxa,Sedrakian:2018ydt,Fortin:2021sst}; magnetar profiles are not taken into account, only simple volume multiplications are performed; no hard X-ray spectra from plausible standard mechanisms are considered \cite{Thompson:2004yg,Heyl:2005an,Beloborodov:2006qh,Baring:2006hi,Lyubarsky:2007gj,Beloborodov:2012ug}, the full hard X-ray spectra are derived from ALPs; \textit{etc.}  With this short note, our primary goal is to introduce the idea and study with simple estimates if such a scenario is plausible, leading us to conclude somewhat positively.


\section{Magnetars}\label{SMag}

The three chosen magnetars and some of their characteristics are shown in Table~\ref{TabMag} \cite{Olausen:2013bpa}.\footnote{See the \href{http://www.physics.mcgill.ca/~pulsar/magnetar/main.html}{McGill Online Magnetar Catalog}.}  They were selected for their very high X-ray luminosities in the soft X-ray range $L_X$ and substantial ages (which exacerbate the age/temperature problem) as well as their large surface magnetic fields $B_0$ (for optimal ALP-to-photon conversion).  Their ages are the characteristic ages $\tau_c=\frac{P}{2\dot{P}}$ obtained from their periods $P$ and their time derivatives $\dot{P}$.  The remaining columns show their distances from Earth $D$ and their surface temperatures $T_s$.
\begin{table}[t]
\centering
\begin{tabular}{|c|ccccc|}
\hline
\multirow{2}{*}{Magnetar} & $L_X$ & $B_0$ & $\tau_c$ & $D$ & $T_s$\\
 & $10^{33}\,\text{erg}\cdot\text{s}^{-1}$ & $10^{14}\,\text{G}$ & $\text{kyr}$ & $\text{kpc}$ & $\text{keV}$\\\hline
SGR 1806-20 & 163 & 20 & 0.24 & 8.7 & 0.55\\
SGR 1900+14 & 90 & 7.0 & 0.90 & 12.5 & 0.47\\
4U 0142+61 & 105 & 1.3 & 68 & 3.6 & 0.410\\
\hline
\end{tabular}
\caption{Relevant characteristics of magnetars studied in our analysis.}
\label{TabMag}
\end{table}

Magnetar core temperatures (without ALP emission) can be estimated from their X-ray luminosities \cite{Beloborodov:2016mmx} and their surface temperatures \cite{Potekhin:2006iw}.  For the three magnetars of interest here, their core temperatures are shown in Table~\ref{TabTcore}.  Considering that the maximal thermal energy \cite{Page:2004fy} and the magnetic energy \cite{Beloborodov:2016mmx} stored in magnetars are approximately
\eqn{E_\text{th}\approx\left(10^{48}\,\text{erg}\right)\left(\frac{T}{10^9\,\text{K}}\right)^2,\qquad\qquad E_\text{mag}\approx\left(10^{43}\,\text{erg}\right)\left(\frac{B_0}{B_c}\right)^2\left(\frac{r_0}{10\,\text{km}}\right)^3,}[EqE]
it is clear that most of the magnetar energy is stored in thermal energy (here $r_0$ is the magnetar radius).  Moreover, without ALP production, neutrino emission is dominated by the modified URCA process\footnote{As long as the neutron star central density is not too large, otherwise the direct URCA process dominates.} \cite{Friman:1979ecl,Raffelt:1996wa}
\eqn{Q_\nu=\frac{11513(9\pi)^{\frac{1}{3}}f^4C_A^2\cos^2\theta_C}{241920}\frac{G_F^2(m_N)^{\frac{7}{3}}}{(m_\pi)^4}T^8\rho^{\frac{2}{3}}\approx\left(10^{21}\,\text{erg}\cdot\text{cm}^{-3}\cdot\text{s}^{-1}\right)\left(\frac{T}{10^9\,\text{K}}\right)^8\left(\frac{\rho}{\rho_0}\right)^{\frac{2}{3}},}[EqQnu]
where $f\approx1$ is the pion-nucleon coupling constant, $C_A\approx-1.26$ is charge-current axial-vector constant, $\theta_C\approx0.24$ is the Cabbibo angle, $G_F\approx1.17\times10^{-5}\,\text{GeV}^{-2}$ is the Fermi constant, $m_N$ and $m_\pi$ are the neutron and pion masses respectively, and $\rho_0\approx2.8\times10^{14}\,\text{g}\cdot\text{cm}^{-3}$ is the nuclear saturation density.

Therefore the neutrino emissivity \eqref{EqQnu} leads to a depletion of the magnetar thermal energy \eqref{EqE} in timescales of the order of
\eqn{t_\text{\cancel{ALP}}\approx\frac{E_\text{th}}{L_\nu(T_\text{\cancel{ALP}})}\approx\left(0.01\,\text{kyr}\right)\left(\frac{T_\text{\cancel{ALP}}}{10^9\,\text{K}}\right)^{-6}\left(\frac{r_0}{10\,\text{km}}\right)^{-3}\left(\frac{\rho}{\rho_0}\right)^{-\frac{2}{3}},}[EqtwoALP]
in tension with the characteristic ages $\tau_c$ shown in Table~\ref{TabMag} for core temperatures without ALP emission.  Here $L_\nu(T_\text{\cancel{ALP}})=VQ_\nu$ is the neutrino luminosity with $V$ the magnetar volume.  We dub this observation the magnetar age/temperature problem.
\begin{table}[t]
\centering
\begin{tabular}{|c|cc|}
\hline
\multirow{2}{*}{Magnetar} & $T_\text{\cancel{ALP}}(T_s)$ & $T_\text{\cancel{ALP}}(L_X)$\\
 & $10^8\,\text{K}$ & $10^8\,\text{K}$\\\hline
SGR 1806-20 & 16.7 & 7.49\\
SGR 1900+14 & 11.4 & 5.30\\
4U 0142+61 & 9.08 & 5.77\\
\hline
\end{tabular}
\caption{Magnetar core temperatures without ALPs from surface temperatures [$T_\text{\cancel{ALP}}(T_s)$] and X-ray luminosities [$T_\text{\cancel{ALP}}(L_X)$].}
\label{TabTcore}
\end{table}

Since ALP emissivity from nucleon-nucleon collisions has a softer core temperature dependence compared to neutrino emission \{$T^6$ for ALPs [see \eqref{EqQa} below] compared to $T^8$ for neutrinos \eqref{EqQnu}\}, that ALPs can convert to photons in the magnetosphere, and that ALP emission peaks in the X-ray range, it is natural to investigate if ALP production can explain (part of) magnetar hard X-ray spectra while allowing for smaller core temperatures and thus alleviate the age/temperature problem.


\section{Axion-like Particles}\label{SALPs}

We now consider the multiple effects of ALPs on magnetars.  First, for $a$ the ALP field, $F_{\mu\nu}$ and $\tilde{F}^{\mu\nu}$ the electromagnetic tensor field and its dual, and $N$ the nucleon field, the SM lagrangian is modified to include
\eqn{\mathcal{L}\supset\frac{1}{2}\partial^\mu a\partial_\mu a-\frac{1}{2}m_a^2a^2-\frac{1}{4}g_{a\gamma}aF_{\mu\nu}\tilde{F}^{\mu\nu}+g_{aN}(\partial_{\mu}a)\bar{N}\gamma^{\mu}\gamma_5N,}[EqL]
where $m_a$ is the ALP mass, $g_{a\gamma}$ is the ALP-photon coupling, and $g_{aN}$ is the ALP-nucleon coupling.  In our scenario, the ALP-nucleon coupling is the dominant coupling for ALP production through nucleon-nucleon collisions while the ALP-photon coupling leads to ALP-to-photon conversion in the magnetosphere.


\subsection{Production}\label{SSProduction}

The dominant mechanism for ALP production in magnetars is ALP bremsstrahlung through nucleon-nucleon collisions, which can be computed from the ALP-nucleon interaction term of \eqref{EqL}.  For degenerate neutron matter encountered in magnetars, the ALP emissivity is obtained from the process $n+n\to n+n+a$ (no exotic processes like superfluidity are considered) in the one-pion-exchange approximation \cite{Iwamoto:1984ir,Raffelt:1996wa},
\eqn{Q_a=\frac{31f^4}{315(9\pi)^{\frac{1}{3}}}\frac{(m_N)^{\frac{11}{3}}}{(m_\pi)^4}g_{aN}^2T^6\rho^{\frac{1}{3}}\approx\left(10^{19}\,\text{erg}\cdot\text{cm}^{-3}\cdot\text{s}^{-1}\right)\left(\frac{g_{aN}}{10^{-10}\,\text{GeV}^{-1}}\right)^2\left(\frac{T}{10^9\,\text{K}}\right)^6\left(\frac{\rho}{\rho_0}\right)^{\frac{1}{3}}.}[EqQa]
It is easily computed from the differential ALP emissivity, which leads to the following differential ALP luminosity \cite{Iwamoto:1984ir,Raffelt:1996wa}
\eqna{
\frac{dL_a(x)}{dx}&=\frac{f^4}{3^{\frac{5}{3}}\pi^{\frac{16}{3}}}\frac{(m_N)^{\frac{11}{3}}}{(m_\pi)^4}g_{aN}^2T^6r_0^3\rho^{\frac{1}{3}}\frac{x^3(x^2+4\pi^2)e^{-x}}{1-e^{-x}}\\
&\approx\left(1.18\times10^{35}\,\text{erg}\cdot\text{s}^{-1}\right)\left(\frac{g_{aN}}{10^{-10}\,\text{GeV}^{-1}}\right)^2\left(\frac{T}{10^9\,\text{K}}\right)^6\left(\frac{r_0}{10\,\text{km}}\right)^3\left(\frac{\rho}{\rho_0}\right)^{\frac{1}{3}}\frac{x^3(x^2+4\pi^2)e^{-x}}{1-e^{-x}},
}[EqLa]
where $x=\omega/(k_BT)$ with $\omega$ the ALP energy and $k_B$ the Boltzmann constant.  The differential ALP luminosity \eqref{EqLa} is computed from the differential ALP emissivity by assuming a constant magnetar profile.  Hence, it is straightforward to derive the differential ALP emissivity from \eqref{EqLa}, and then integrate over the ALP energy to generate \eqref{EqQa}.


\subsection{Conversion}\label{SSConversion}

Once ALPs are produced, they escape the magnetars unhindered due to their large mean free paths \cite{Raffelt:1987im,Lai:2006af}.  They are then allowed to oscillate into photons in the magnetosphere of the magnetars with the help of the ALP-photon coupling in \eqref{EqL}.  Since most of the ALP-to-photon conversion occurs extremely far ($\sim1000r_0$) from the magnetar surface, the magnetic field at the conversion radius is mostly dipolar and subcritical.  Following \cite{Raffelt:1987im,Fortin:2018ehg,Fortin:2018aom}, inasmuch as it is much smaller than one, the ALP-to-photon conversion probability at infinity is given by
\eqn{P_{a\to\gamma}(x,\theta)=\left(\frac{1}{2}g_{a\gamma}r_0B_0\sin\theta\right)^2\left|\int_1^\infty dt\,\frac{1}{t^3}\exp\left[i\left(-\frac{\zeta t}{x}+\frac{\xi x\sin^2\theta}{t^5}\right)\right]\right|^2,}[EqPatogamma]
where
\eqn{\zeta=\frac{m_a^2r_0}{2k_BT},\qquad\qquad\xi=\frac{7\alpha r_0}{450\pi}\left(\frac{B_0}{B_c}\right)^2k_BT,}[Eqzetaxi]
and $\theta$ is the angle between the direction of propagation of the ALP/photon and the magnetic field.  Since ALPs should be produced isotropically by the magnetar core, it is natural to introduce the averaged ALP-to-photon conversion probability, given by
\eqn{\bar{P}_{a\to\gamma}(x)=\frac{1}{2\pi}\int_0^{2\pi}d\theta\,P_{a\to\gamma}(x,\theta),}[EqPavg]
from \eqref{EqPatogamma}.

Before proceeding to the differential ALP-to-photon luminosity, obtained by multiplying the ALP luminosity and the averaged ALP-to-photon conversion probability, it is convenient to discuss an approximation to the ALP-to-photon conversion probability, valid in the limit $(\zeta/x)^{5/6}(\xi x)^{1/6}\lesssim0.34$ \cite{Fortin:2018aom}.  In this limit, we have
\eqn{P_{a\to\gamma}(x,\theta)=\left(\frac{1}{2}g_{a\gamma}r_0B_0\sin\theta\right)^2\frac{\Gamma\!\left(\frac{2}{5}\right)^2}{25(\xi x\sin^2\theta)^\frac{4}{5}},\qquad\qquad\bar{P}_{a\to\gamma}(x)=\left(\frac{1}{2}g_{a\gamma}r_0B_0\right)^2\frac{2^\frac{3}{5}\Gamma\!\left(\frac{2}{5}\right)^3}{5\Gamma\!\left(\frac{1}{5}\right)^2(\xi x)^\frac{4}{5}},}[EqPapprox]
from which it is possible to find an analytic approximation to the differential ALP-to-photon luminosity.


\subsection{Photon Spectrum from ALPs}\label{SSSpectrum}

We are now ready to determine the differential ALP-to-photon luminosity in the scenario where photon emission in the hard X-ray range is negligible.  Indeed, \eqref{EqLa} and \eqref{EqPavg} can be combined to obtain the ALP-to-photon luminosity as in
\eqn{\frac{dL_{a\to\gamma}(x)}{dx}=\frac{dL_a(x)}{dx}\bar{P}_{a\to\gamma}(x).}[EqLatogamma]
To better compare with observations, it is actually more convenient to introduce
\eqn{\nu F_\nu(\omega)=\omega^2\frac{1}{4\pi D^2}\frac{1}{\omega}\frac{dL_a(\omega)}{d\omega},}[EqnuFnu]
instead of \eqref{EqLatogamma}, which is used by various collaborations when displaying their spectra.

From our analysis, the theoretical photon spectrum \eqref{EqnuFnu} becomes
\eqna{
\nu F_\nu(\omega)&=\frac{5^{\frac{3}{5}}\Gamma\!\left(\frac{2}{5}\right)^3f^4}{2^\frac{21}{5}3^{\frac{1}{15}}7^{\frac{4}{5}}\pi^{\frac{19}{3}}\Gamma\!\left(\frac{1}{5}\right)^2\alpha^{\frac{8}{5}}}\frac{(m_e)^{\frac{16}{5}}(m_N)^{\frac{11}{3}}}{(m_\pi)^4}\frac{g_{aN}^2g_{a\gamma}^2T^{\frac{26}{5}}r_0^{\frac{21}{5}}B_0^{\frac{2}{5}}\rho^{\frac{1}{3}}}{D^2}I(\zeta/x,\xi x)\frac{x^\frac{16}{5}(x^2+4\pi^2)e^{-x}}{1-e^{-x}}\\
&\approx\left(5.09\times10^{-4}\,\text{keV}^2\cdot\text{photons}\cdot\text{cm}^{-2}\cdot\text{s}^{-1}\cdot\text{keV}^{-1}\right)\left(\frac{g_{aN}}{10^{-10}\,\text{GeV}^{-1}}\right)^2\left(\frac{g_{a\gamma}}{10^{-10}\,\text{GeV}^{-1}}\right)^2\\
&\phantom{=}\times\left(\frac{T}{10^9\,\text{K}}\right)^\frac{26}{5}\left(\frac{r_0}{10\,\text{km}}\right)^\frac{21}{5}\left(\frac{B_0}{B_c}\right)^\frac{2}{5}\left(\frac{\text{kpc}}{D}\right)^2\left(\frac{\rho}{\rho_0}\right)^{\frac{1}{3}}I(\zeta/x,\xi x)\frac{x^\frac{16}{5}(x^2+4\pi^2)e^{-x}}{1-e^{-x}},
}[EqnuFnuth]
where $I(\zeta/x,\xi x)$ is a function of the ALP and magnetar parameters \eqref{Eqzetaxi} as well as the ALP energy divided by the temperature,
\eqn{I(\zeta/x,\xi x)=\frac{5\Gamma\!\left(\frac{1}{5}\right)^2}{2^\frac{3}{5}\pi\Gamma\!\left(\frac{2}{5}\right)^3}(\xi x)^\frac{4}{5}\int_0^1du\,\sqrt{\frac{u}{1-u}}\left|\int_0^1dt\,t\exp\left[i\left(-\frac{\zeta}{xt}+\xi xut^5\right)\right]\right|^2.}[EqI]
The dimensionless integral \eqref{EqI} corresponds to the dimensionless averaging between the ALP/photon direction of propagation and the magnetic field (with $u=\sin^2\theta$) of the ALP-to-photon conversion \eqref{EqPavg} [with the change of variable $t\to1/t$ in \eqref{EqPatogamma}], normalized to the approximation \eqref{EqPapprox}.  Hence, the integral \eqref{EqI} in the theoretical photon spectrum \eqref{EqnuFnuth} is of order one in the appropriate limit, which is (mostly) the case for the magnetars under consideration.  Therefore, setting $I(\zeta/x,\xi x)\to1$ in \eqref{EqnuFnuth} is a fair analytic result for the theoretical photon spectrum as long as $(\zeta/x)^{5/6}(\xi x)^{1/6}\lesssim0.34$.


\section{Results}\label{SResults}

To proceed, we use the datasets from Enoto \textit{et al.} \cite{Enoto:2017dox} and Younes \textit{et al.} \cite{Younes:2017sme} for SGR 1806-20, the dataset from Tamba \textit{et al.} \cite{Tamba:2019als} for SGR 1900+14, as well as the two datasets from Enoto \textit{et al.} \cite{Enoto:2011aa,Enoto:2017dox} and one dataset from Wang \textit{et al.} \cite{Wang:2013pka} for 4U 0142+61.  We then perform $\chi^2$ fits following
\eqn{\chi^2=\sum_{i=1}^N\frac{(O_i-E_i)^2}{\sigma_i^2},}
for datasets with $N$ data points, where $O_i\pm\sigma_i$ is an observed value and its error bar while $E_i$ is a value computed from the model, mainly \eqref{EqnuFnuth}.

\begin{figure}[!t]
\centering
\resizebox{15cm}{!}{
\includegraphics{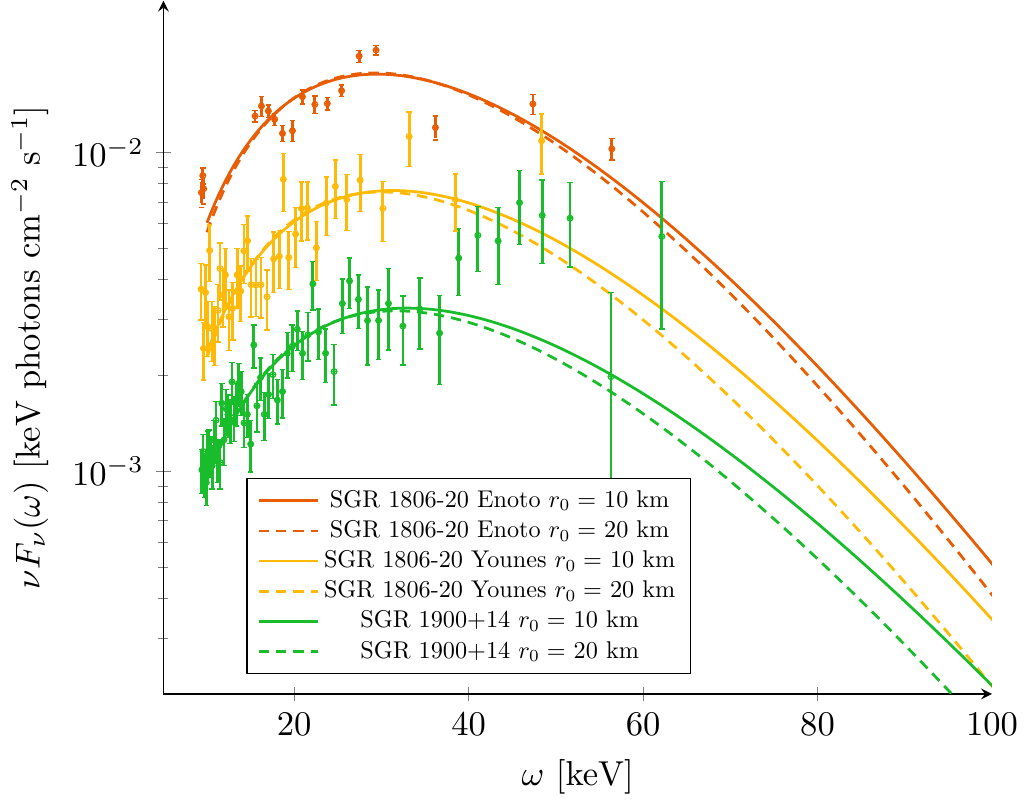}
\hspace{2cm}
\includegraphics{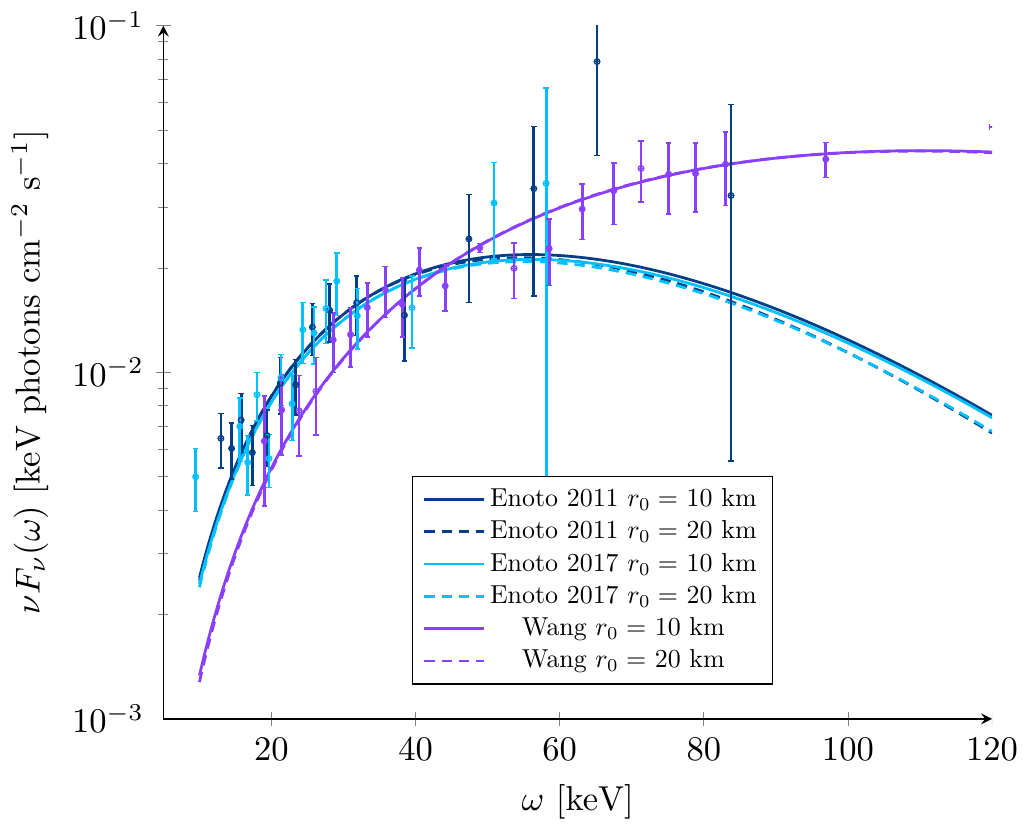}
}
\caption{Comparison between the observed (with different datasets) and theoretical spectra for the three magnetars of Table~\ref{TabMag}.  The data points correspond to the different datasets while the solid and dashed lines correspond to magnetar radii of $10\,\text{km}$ and $20\,\text{km}$, respectively.}
\label{FignuFnu}
\end{figure}

\begin{table}[!t]
\centering
\begin{tabular}{|c|ccc|ccc|}
\hline
\multirow{3}{*}{Magnetar} & \multicolumn{3}{c|}{$r_0=10\,\text{km}$} & \multicolumn{3}{c|}{$r_0=20\,\text{km}$}\\
 & $\chi^2/n_d$ & $T$ & $|g_{aN}g_{a\gamma}|$ & $\chi^2/n_d$ & $T$ & $|g_{aN}g_{a\gamma}|$\\
 & $-$ & $10^8\,\text{K}$ & $(10^{-10}\,\text{GeV}^{-1})^2$ & $-$ & $10^8\,\text{K}$ & $(10^{-10}\,\text{GeV}^{-1})^2$\\\hline
\underline{SGR 1806-20} & & & & & &\\
Enoto \textit{et al.} & 10.8 & 0.94 & 1370 & 12.3 & 0.90 & 369\\
Younes \textit{et al.} & 1.3 & 0.99 & 767 & 1.4 & 0.92 & 224\\\hline
\underline{SGR 1900+14} & & & & & &\\
Tamba \textit{et al.} & 1.4 & 1.09 & 708 & 1.5 & 1.03 & 196\\\hline
\underline{4U 0142+61} & & & & & &\\
Enoto \textit{et al.} (2011) & 1.2 & 1.80 & 192 & 1.2 & 1.75 & 49\\
Enoto \textit{et al.} (2017) & 1.7 & 1.81 & 187 & 1.7 & 1.77 & 47\\
Wang \textit{et al.} & 0.6 & 3.56 & 46 & 0.6 & 3.53 & 11\\
\hline
\end{tabular}
\caption{Best-fit core temperatures and ALP coupling constants for the three magnetars of Table~\ref{TabMag} assuming their hard X-ray spectra are generated solely by ALP-to-photon conversion.  The number of degrees of freedom is defined as $n_d=N-n_p$ where $N$ is the number of data points per dataset and $n_p$ is the number of free parameters (here $n_p=2$).}
\label{TabBestFit}
\end{table}
We note that we leave the product of ALP coupling constants $|g_{aN}g_{a\gamma}|$ and the core temperature $T$ free in our fits.  For the other quantities, we use the magnetar parameters of Table~\ref{TabMag} and fix the nuclear density to the nuclear saturation density, $\rho\to\rho_0$, and the ALP mass to $m_a=10^{-5}\,\text{eV}$.\footnote{We note that any ALP mass---which appears only in the dimensionless integral \eqref{EqI} through its dependence on $\zeta$ \eqref{Eqzetaxi}---below $10^{-5}\,\text{eV}$ leads to the same result due to the properties of the integral \cite{Fortin:2018ehg}.}  Moreover, since magnetar radii are hard to estimate, we perform the fits for two different values of radii, $10\,\text{km}$ and $20\,\text{km}$.  Our results are shown in Table~\ref{TabBestFit} as well as Figure~\ref{FignuFnu}.

Table~\ref{TabBestFit} shows that our scenario where ALPs are responsible for the magnetar quiescent X-ray spectra is reasonable since all but one ratios $\chi^2/n_d$ are order one.  Figure~\ref{FignuFnu} clearly demonstrates however that the two datasets for SGR 1806-20, with the worst best-fit dataset by Enoto \textit{et al.}, are somewhat incompatible.  The discrepancy may originate from the different instruments used as well as the different periods they observed the magnetar.  Hence, barring a thorough statistical analysis combining the different datasets for each magnetar, the worst best-fit dataset may be discarded in the following, although it does lead to matching core temperature and ALP coupling constants.  Table~\ref{TabBestFit} also shows a decrease of up to one order of magnitude of the magnetar core temperatures compared to the standard scenario with only neutrinos (depending on the method used in Table~\ref{TabTcore}).  These observations are valid for both magnetar radii.  As expected, the best-fit values of the ALP coupling constants $|g_{aN}g_{a\gamma}|$ are slightly smaller for the larger radius since bigger magnetars emit more ALPs than smaller magnetars for the same emissivity (energy emitted per unit time per unit volume).  Since the dimensionless integral \eqref{EqI} is more or less constant for the two magnetar radii, from \eqref{EqnuFnuth} the ratio of the ALP coupling constants with $r_0=20\,\text{km}$ to the ALP coupling constants with $r_0=10\,\text{km}$ is approximately $2^{\frac{21}{10}}\approx4.3$, in agreement with Table~\ref{TabBestFit}.  As argued below, the best-fit ALP coupling constants are however somewhat large.\footnote{We point out in passing that the best-fit values for $|g_{aN}g_{a\gamma}|$ cannot be distributed arbitrarily between $g_{aN}$ and $g_{a\gamma}$.  Indeed, the ALP-photon coupling $g_{a\gamma}$ must be small enough to ensure the consistency of the ALP-to-photon conversion probability \eqref{EqPatogamma} which cannot become too large (for example, it cannot be larger than one).}

Now that we have established that our scenario is plausible, we can compare it to existing bounds and investigate its implications on the age/temperature problem.  For instance, to alleviate the age/temperature problem, it is necessary that the sum of the luminosities in ALPs and neutrinos at the best-fit core temperature is smaller than the luminosity in neutrinos only with core temperature without ALPs.  This constraint,
\eqn{L_\nu(T)+L_a(T)\leq L_\nu(T_\text{\cancel{ALP}}),}
implies that
\eqna{
g_{aN}\leq g_{aN,L}(T)&=\frac{3^{\frac{1}{6}}29^{\frac{1}{2}}397^{\frac{1}{2}}\pi^{\frac{1}{3}}C_A\cos\theta_C}{2^431^{\frac{1}{2}}}\frac{G_F}{(m_N)^{\frac{2}{3}}}\rho^{\frac{1}{6}}T\sqrt{\left(\frac{T_\text{\cancel{ALP}}}{T}\right)^8-1}\\
&\approx\left(8.89\times10^{-10}\,\text{GeV}^{-1}\right)\left(\frac{\rho}{\rho_0}\right)^{\frac{1}{6}}\left(\frac{T}{10^9\,\text{K}}\right)\sqrt{\left(\frac{T_\text{\cancel{ALP}}}{T}\right)^8-1}.
}[EqLBound]
As such, the bound \eqref{EqLBound} is not convenient since it cannot be compared directly with the best-fit values of Table~\ref{TabBestFit}.  There are however extra bounds originating from other observations.  We use here the supernova SN1987a bound \cite{Raffelt:1996wa} on the ALP-nucleon coupling and the CAST bound \cite{Anastassopoulos:2017ftl} on the ALP-photon coupling,
\eqn{g_{aN}\leq g_\text{SN1987a}=8.0\times10^{-10}\,\text{GeV}^{-1},\qquad\qquad g_{a\gamma}\leq g_\text{CAST}=0.66\times10^{-10}\,\text{GeV}^{-1},}[EqgBounds]
to produce two useful bounds.  Indeed, combining the bounds \eqref{EqLBound} and \eqref{EqgBounds} thus leads to the two following bounds
\eqn{
\begin{gathered}
\frac{|g_{aN}g_{a\gamma}|}{(10^{-10}\,\text{GeV}^{-1})^2}\leq5.9\left(\frac{\rho}{\rho_0}\right)^{\frac{1}{6}}\left(\frac{T}{10^9\,\text{K}}\right)\sqrt{\left(\frac{T_\text{\cancel{ALP}}}{T}\right)^8-1},\\
\frac{|g_{aN}g_{a\gamma}|}{(10^{-10}\,\text{GeV}^{-1})^2}\leq5.3,
\end{gathered}
}[EqggBound]
on the product of the ALP coupling constants, which can be conveniently compared to the best-fit values obtained in Table~\ref{TabBestFit}.

\begin{figure}[!t]
\begin{minipage}{0.5\linewidth}
\centering
\subfloat[]{\label{FigContour-SGR1806-20}\includegraphics[scale=0.65]{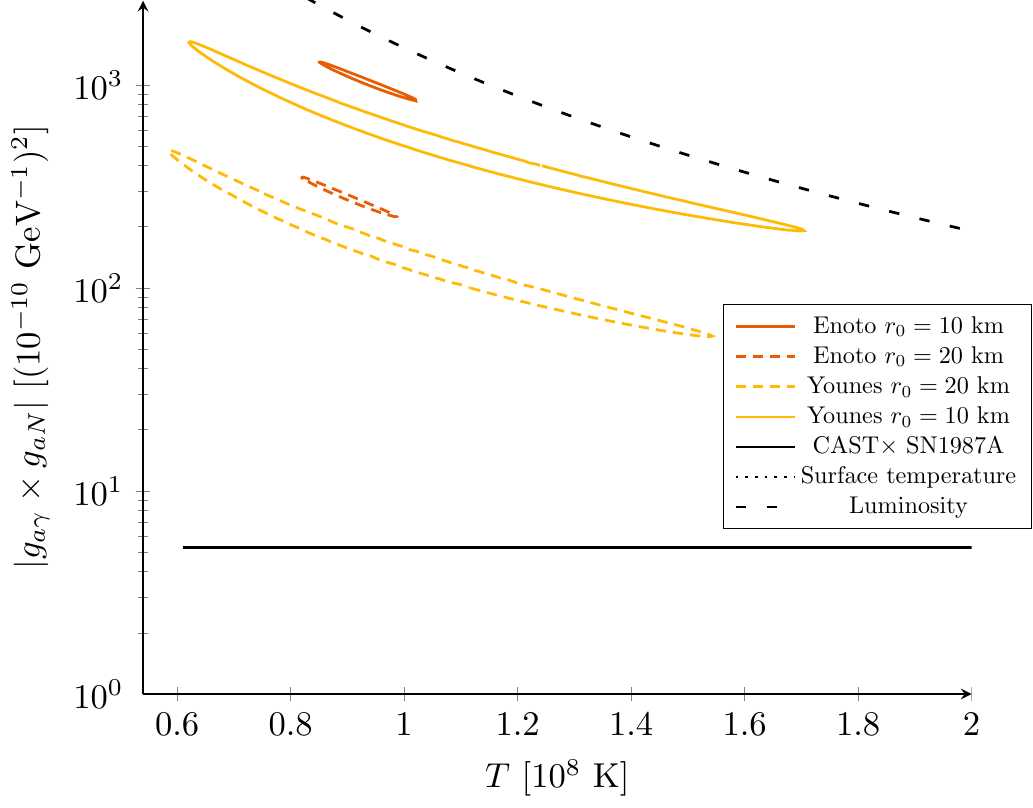}}
\end{minipage}
\begin{minipage}{0.5\linewidth}
\centering
\subfloat[]{\label{FigContour-SGR1900+14}\includegraphics[scale=0.65]{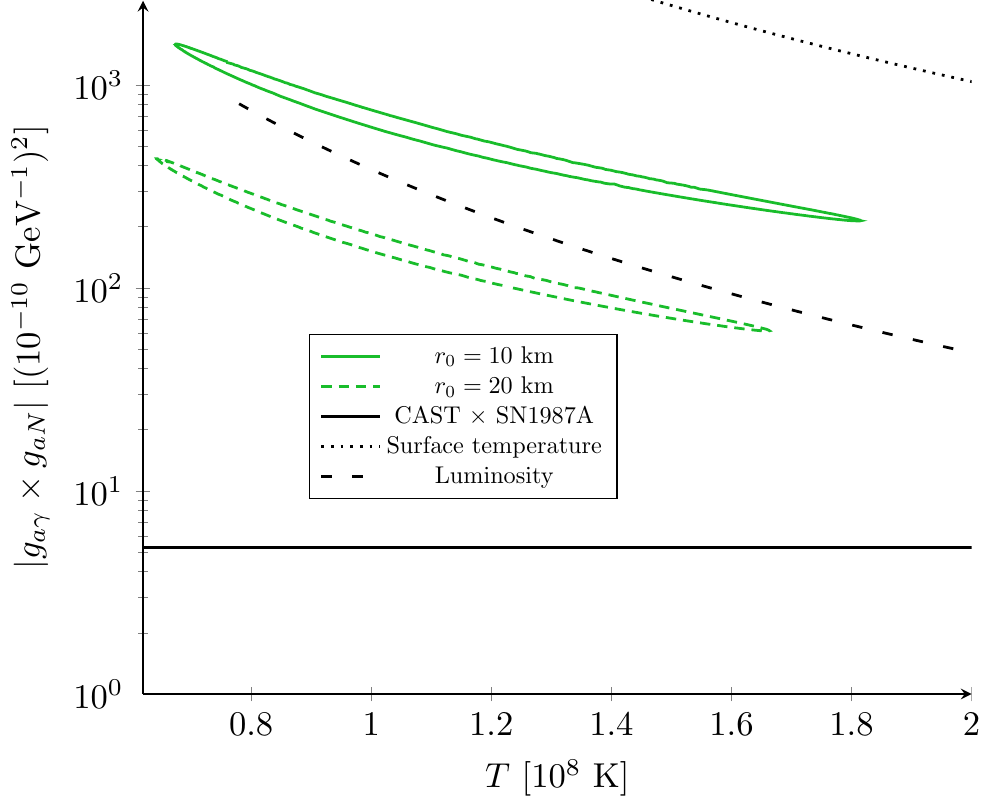}}
\end{minipage}\par\medskip
\begin{minipage}{0.5\linewidth}
\centering
\subfloat[]{\label{FigContour-4U0142+61}\includegraphics[scale=0.65]{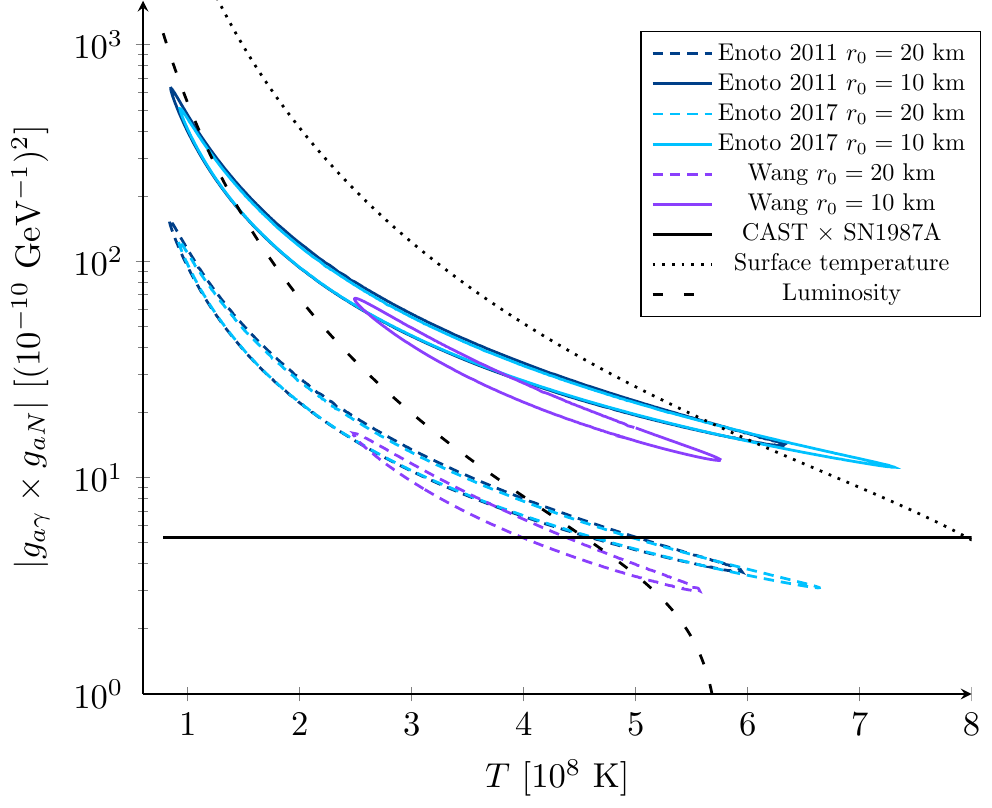}}
\end{minipage}
\begin{minipage}{0.5\linewidth}
\centering
\subfloat[]{\label{FigContour-All}\includegraphics[scale=0.65]{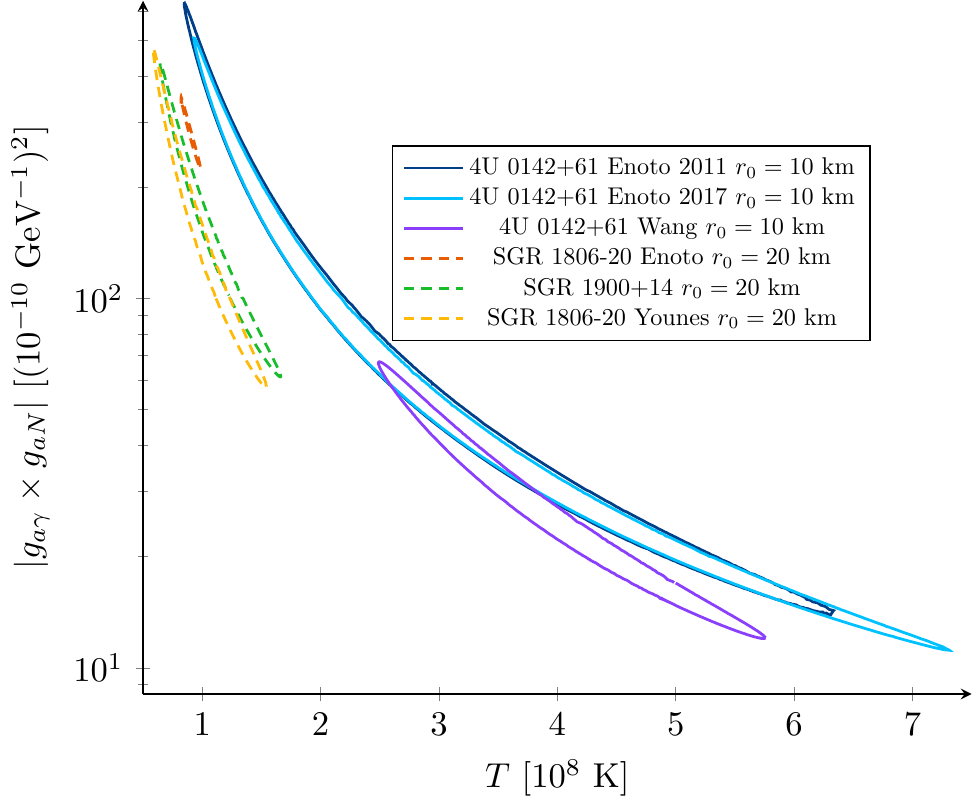}}
\end{minipage}
\caption{$90\%$ confidence-level contours for $r_0=10\,\text{km}$ (solid colored lines) and $r_0=20\,\text{km}$ (dashed colored lines) as well as the bounds \eqref{EqggBound} (solid, dashed, and dotted black lines) for the three magnetars of Table~\ref{TabMag} (Figure~\ref{FigContour-SGR1806-20} for SGR 1806-20, Figure~\ref{FigContour-SGR1900+14} for SGR 1900+14, and Figure~\ref{FigContour-4U0142+61} for 4U 0142+61).  Comparison of several $90\%$ confidence-level contours for different magnetars with different radii in the magnetar core temperature/ALP coupling constants plane (Figure~\ref{FigContour-All}).}
\label{FigContour}
\end{figure}
To proceed, we compute the $90\%$ confidence-level contours in the core temperature/ALP coupling constants plane for the two magnetar radii and superimpose the two bounds \eqref{EqggBound}.  The results are shown in Figures~\ref{FigContour-SGR1806-20},~\ref{FigContour-SGR1900+14}, and~\ref{FigContour-4U0142+61}.  At first glance, it is clear that although including ALP emission allows for smaller core temperatures, the best-fit scenarios are still in tension with the bounds of \eqref{EqggBound}, especially the combined bound from SN1987a and CAST (the solid black lines).  As mentioned above, the larger magnetar radius $r_0=20\,\text{km}$ leads to less tension with the two bounds \eqref{EqggBound}, although the problem persists.  With respect to the luminosity bound on the ALP coupling constants (dashed and dotted black lines), the small value of the ALP-photon coupling from CAST leads to strong bounds although the core temperatures with ALP emission are much smaller than core temperatures without ALP emission.  Indeed, starting from the luminosity bound on the ALP-nucleon coupling \eqref{EqLBound}, the age/temperature problem can still be addressed since magnetar lifetimes can be estimated from
\eqna{
t\approx\frac{E_\text{th}}{L_\nu(T)+L_a(T)}&\approx\left(0.01\,\text{kyr}\right)\left(\frac{T}{10^9\,\text{K}}\right)^{-6}\left(\frac{r_0}{10\,\text{km}}\right)^{-3}\left(\frac{\rho}{\rho_0}\right)^{-\frac{2}{3}}\\
&\phantom{\approx}\qquad\times\left[1+0.01\left(\frac{g_{aN}}{10^{-10}\,\text{GeV}^{-1}}\right)^2\left(\frac{T}{10^9\,\text{K}}\right)^{-2}\left(\frac{\rho}{\rho_0}\right)^{-\frac{1}{3}}\right]^{-1}.
}[EqtwALP]
Even though \eqref{EqtwoALP} and \eqref{EqtwALP} are quite alike, with the (smaller) best-fit temperatures of Table~\ref{TabBestFit}, it is now possible to generate lifetimes in agreement with the observed magnetar ages of Table~\ref{TabMag} for two magnetars and alleviate the issue by at least two orders of magnitude for the third magnetar, as shown in the last column of Table~\ref{TabLife}.
\begin{table}[t]
\centering
\begin{tabular}{|c|c|cc|c|}
\hline
\multirow{2}{*}{Magnetar} & $\tau_c$ & $t_\text{\cancel{ALP}}(T_s)$ & $t_\text{\cancel{ALP}}(L_X)$ & $t$\\
 & $\text{kyr}$ & $\text{kyr}$ & $\text{kyr}$ & $\text{kyr}$\\\hline
SGR 1806-20 & $2\times10^{-1}$ & $4\times10^{-4}$ & $5\times10^{-2}$ & $1\times10^2$\\
SGR 1900+14 & $9\times10^{-1}$ & $4\times10^{-3}$ & $4\times10^{-1}$ & $8\times10^1$\\
4U 0142+61 & $7\times10^1$ & $2\times10^{-2}$ & $2\times10^{-1}$ & $1\times10^1$\\
\hline
\end{tabular}
\caption{Comparison between approximate magnetar ages and estimated lifetimes, both without ALPs \eqref{EqtwoALP} (for core temperatures estimated by the surface temperature or the luminosity) and with ALPs \eqref{EqtwALP} (for best, hence minimal, core temperatures found in Table~\ref{TabBestFit}).  In all cases, $r=10\,\text{km}$ and $\rho=\rho_0$, while $g_{aN}=g_\text{SN1987a}$ \eqref{EqgBounds}.}
\label{TabLife}
\end{table}
These estimates are computed assuming the upper bound on the ALP-nucleon coupling constant \eqref{EqgBounds}.  It is worth pointing out that smaller values for the ALP-nucleon coupling constant would lead to longer lifetimes.  Therefore a plausible explanation of the age/temperature problem from this scenario for the third magnetar is not ruled out.

Moreover, since there are plausible non-exotic mechanisms to generate (part of) the hard X-ray magnetar spectra \cite{Thompson:2004yg,Heyl:2005an,Beloborodov:2006qh,Baring:2006hi,Lyubarsky:2007gj,Beloborodov:2012ug}, it would be of interest to consider one or more of these mechanisms with ALP emission to determine the best-fit parameters.  Although they cannot solve the magnetar age/temperature problem by themselves, the interplay of these different mechanisms with ALP production/conversion could allow ALP emission to lower the magnetar core temperatures and solve the age/temperature problem while keeping the ALP coupling constants consistent with the SN1987a and CAST bounds.  Indeed, since such mechanisms rely on the strength of the magnetic field and the chosen magnetars have extreme magnetic fields (for ALP-photon conversion purposes), considering both type of mechanisms could alleviate the issue, although a thorough analysis of such scenarios is outside the scope of this short note.\footnote{Besides including these non-exotic backgrounds, such an analysis should relax the simplifications mentioned above by considering: protons, the soft-radiation approximation, superfluidity, \textit{etc.}}

Finally, for completeness we superimpose in Figure~\ref{FigContour-All} the $90\%$ confidence-level contours of the different magnetars.  Indeed, although the magnetars can have different core temperatures and radii, our scenario forces the best-fit ALP coupling constants to be the same.  Figure~\ref{FigContour-All} shows that for different magnetar radii, the ALP coupling constants are somewhat compatible between different magnetars.  It would be interesting to see if the agreement can be improved by including hard X-ray emission from non-exotic mechanisms as discussed above.  In addition, better estimates of magnetar radii would be welcomed to investigate this possibility in more detail.


\section{Conclusion}\label{SConclusion}

Magnetars are extreme stellar objects that can be used as natural laboratories to investigate the existence and would-be properties of extremely light and weakly-interacting exotic particles.  In the context of this short paper, magnetars were used as a source of axion-like-particles (ALPs) through the ALP-nucleon coupling which then converted into photons in the magnetosphere through the ALP-photon coupling, leading to a new component of the photon luminosity.

The main motivation centered on the difficulty explaining the hard X-ray magnetar spectra from known mechanisms and the large magnetar core temperatures necessary to reproduce the soft X-ray magnetar spectra, leading to extreme cooling through neutrino emission in contradiction to the ages of magnetars as determined from their periods and their derivatives.  At first glance it would seem that including an extra source of energy depletion through ALP emission would exacerbate this magnetar age/temperature problem, but the subsequent conversion of ALPs to photons can lead to the opposite effect by contributing to the X-ray photon spectra and by consequently decreasing the magnetar core temperatures.  Indeed, since neutrino emission in magnetars is dominated by the modified URCA process with $T^8$ temperature dependence while ALP emission in magnetars is dominated by nucleon-nucleon bremsstrahlung with $T^6$ temperature dependence, any change in core temperature has a steep effect on allowed magnetar lifetimes.

In this paper we thus assumed a simplified scenario where the full hard X-ray magnetar spectra are generated by ALP emission and conversion.  We showed that the scenario does indeed allow for smaller core temperatures (by up to one order of magnitude) while explaining the observed data for three magnetars with particular properties (large luminosities and surface magnetic fields).  We also showed that the best-fit ALP coupling constants are however in great tension with bounds coming from other stellar objects, mainly supernovae (SN1987a) and main-sequence stars (sun), although they allowed for a somewhat consistent picture when compared among the three magnetars.  Apart from a more precise treatment of ALP emission, improvements on this scenario might be made by including ALP emission and conversion to plausible known mechanisms proposed in the literature.  We hope to return to this idea in subsequent work.


\ack{
We would like to thank Kuver Sinha for comments on the manuscript.  JFF is supported by NSERC.  The work of MG was supported in part by NSERC.
}


\bibliography{MagnetarHardXRayAxion}


\end{document}